\documentclass{aa}  
\usepackage{graphicx}
\usepackage{graphics}
\usepackage{txfonts}
\usepackage{amssymb}
\usepackage{lscape}
\usepackage{epsfig}
\usepackage{natbib}
\usepackage{multirow}

\usepackage{xcolor}

\usepackage{graphicx}	
\usepackage{amsmath}	
\usepackage{amssymb}	

\usepackage{graphicx}
\usepackage{array}
\usepackage{amssymb}
\usepackage{float}
\usepackage{color}
\usepackage[utf8]{inputenc}  
\usepackage{amsmath}  
\usepackage{comment}
\usepackage{footnote}
\usepackage{longtable}
\usepackage[figuresright]{rotating}
\usepackage{booktabs}
\usepackage[T1]{fontenc}
\usepackage{footnote}
\usepackage{pdfpages}
\usepackage{epstopdf}
\usepackage{graphicx}   
\usepackage{amsmath}    
\usepackage{amssymb}    
\usepackage{graphicx}
\usepackage{array}
\usepackage{amssymb}
\usepackage{float}
\usepackage{color}
\usepackage[utf8]{inputenc}  
\usepackage{amsmath}  
\usepackage{comment}
\usepackage{footnote}
\usepackage{longtable}
\usepackage[figuresright]{rotating}
\usepackage{booktabs}
\usepackage[T1]{fontenc}
\usepackage{footnote}
\usepackage{pdfpages}
\usepackage{epstopdf}
\bibpunct{(}{)}{;}{a}{}{,} 
\begin{document}

   \title{Long term $\gamma$-ray variability of blazars}


   \author{Bhoomika Rajput
          \inst{1}
          \and
          C. S. Stalin 
          \and
          \inst{1}
          \and
          Suvendu Rakshit 
          \inst{2}
          }

   \institute{$^1$ Indian Institute of Astrophysics, Koramangala, Bangalore 560034, India \\
             $^2$ Finnish Centre for Astronomy with ESO (FINCA), 
University of Turku, Quantum, Vesilinnantie 5, 20014, Finland \\
             }

   \date{Received October 10, 2019; accepted December 25, 2019}

 
\abstract
   {
We used the data from the {\it Fermi} 
Gamma-ray Space Telescope to characterise the $\gamma$-ray flux variability of 
blazars on month-like time scales. Our sample consists of 1120 blazars of which 
481 are flat spectrum radio quasars (FSRQs) and 639 are BL Lac objects 
(BL Lacs).  We generated monthly binned light curves of our sample for a period
of approximately nine years from 2008 August to 2017 December and quantified variability by using excess variance ($F_{var}$). On month-like time scales, 371/481 FSRQs  
are variable (~80\%), while only about 50\% (304/639) of BL Lacs 
are variable. This suggests that  FSRQs are more variable than BL Lac objects. We find a mean 
$F_{var}$ of 0.55 $\pm$ 
0.33 and 0.47 $\pm$ 0.29 for FSRQs and BL Lacs respectively. Large
$F_{var}$ in FSRQs is also confirmed from the analysis of the ensemble structure function. By Dividing our 
sample of blazars based on the position of the synchrotron peak in their 
broad-band spectral energy distribution, we find that the low synchrotron 
peaked (LSP) sources have the largest mean $F_{var}$ value of 0.54 $\pm$ 
0.32 while the 
intermediate synchrotron peaked (ISP) and high synchrotron peaked (HSP) 
sources have mean $F_{var}$ values of 0.45 $\pm$ 0.25 and 0.47 $\pm$ 0.33 
respectively. On month-like time scales, we find FSRQs 
to show a high duty cycle (DC) of variability of 66\% relative to BL Lacs
that show a DC of 36\%. We find that both the $F_{var}$ and time scale of variability 
($\tau$) do not correlate with M$_{BH}$. We note that $F_{var}$ is found to be weakly 
correlated with Doppler factor
($\delta$) and $\tau$ is also weakly correlated with $\delta$. Most of the sources in our sample
have $\tau$ of the order of days, which might be related to processes in the jet. We find marginal difference in the 
distribution of $\tau$ between FSRQs and BL Lacs.
   }


\titlerunning{$\gamma$-ray flux variability of AGN}
   \maketitle
%

\section{Introduction}
Flux variability which involves non-periodic changes in flux occurring with different 
amplitudes and time scales is one 
of the defining characteristics of active galactic nuclei (AGN) and it was recognised in these 
objects soon after their discovery about half a century ago \citep{1963ApJ...138...30M}.
Blazars are a peculiar category of radio-loud AGN, with bolometric luminosity
as large as 10$^{48}$ erg s$^{-1}$ or 10$^{14}$ L$_{\odot}$  where their relativistic jets
are pointed close to the line of sight to the observer \citep{1995PASP..107..803U}. They are copious emitters
of high-energy
radiation and show rapid and large amplitude flux variations over the entire accessible
spectral region from low-energy radio to high-energy $\gamma$-rays
\citep{1997ARA&A..35..445U}. They  dominate the extragalactic
$\gamma$-ray sky as revealed by both the Compton Gamma Ray Observatory
\citep{1999ApJS..123...79H} and the {\it Fermi} Gamma Ray Space
Telescope \citep{2019arXiv190210045T}. Blazars comprise both flat spectrum radio quasars (FSRQs) and BL Lacertae objects
(BL Lacs). While FSRQs have
broad emission lines in their optical spectra, BL Lacs have either a featureless optical
spectra or optical spectra with weak (equivalent width $<$ 5 \AA) emission lines. Alternatively,
\cite{2011MNRAS.414.2674G} propose a more physical distinction between FSRQs and BL Lacs which is based on the
luminosity of the broad line region ($L_{BLR}$) relative to the Eddington luminosity ($L_{Edd}$), where
$L_{Edd}$ = 1.38 $\times$ 10$^{38}$ ($M_{BH}/M_{\odot}$) erg sec$^{-1}$, and $M_{BH}$ is the
mass of the black hole.
FSRQs with $L_{BLR}/L_{Edd} > 5 \times 10^{-5}$ are believed to be the
beamed counterparts of the more luminous Fanaroff \& Riley type II (FRII; \citealt{1974MNRAS.167P..31F}) radio sources, while
BL Lacs are the beamed counterparts of the less luminous FRI type radio sources. The broad-band
spectral energy distribution (SED) of blazars in the log $\nu F_{\nu}$ - log $\nu$ representation  has a
two-component structure, with the low-energy component covering the radio to the ultraviolet (UV) and X-ray. The structure is explained by synchrotron emission 
processes and the high-energy component (covering X-ray to $\gamma$-ray), which is explained by inverse Compton emission
processes from relativistic electrons in their  jets. Based on the location of
the peak ($\nu_p$) of the synchrotron emission in their broad-band SED, 
blazars are further divided into low synchrotron peaked blazars with 
$\nu_p < 10^{14}$ Hz, intermediate synchrotron peaked blazars
with $10^{14} Hz \le \nu_p \le 10^{15} Hz,$ and high synchrotron peaked blazars
with $\nu_p > 10^{15} Hz$. The majority of the FSRQs belong to the LSP category,
while a large fraction of HSP sources are BL Lacs.

Since the jets in blazars are aligned close to the observer in the beaming
model, the observed emission ($S_{obs}$) from the jet is Doppler boosted 
relative to what is measured in the co-moving frame of the jet ($S_{int}$) as 
$S_{obs} = S_{int} \delta^q$ \citep{2017RAA....17...66L}
where q = 2 + $\alpha$ for a stationary jet and q = 3 + $\alpha$ for 
a jet with distinct blobs, $\alpha$ is the spectral index
defined as $S_{\nu} \propto \nu^{-\alpha}$, 
$\delta$ is the Doppler factor given by $\delta = 
[\Gamma(1-\beta cos\theta)]^{-1}$,
where $\Gamma$ = $(1 - \beta^2)^{-1/2}$ is the bulk Lorentz factor,  
$\theta$ is the angle between the observer's line of
sight and the jet axis and $\beta = v/c$ is the jet speed. In addition to flux 
enhancement, the observed time scale of variability
is also shortened by a factor $\delta^{-1}$, which is relative to that of the co-moving frame.
These two effects increase our chances of detecting variations in blazars over a range of time
scales and amplitudes. Characterising the minimum time scale of 
variability ($t_{min}$) from
blazar light curves is important as it provides important constraints
on the size of the emitting region in blazar jets via 
$R < c t_{min} \delta (1 + z)^{-1}$. Flux variations  on minute time scales have
been observed in optical, IR and X-ray regimes. Additionally, in high-energy $\gamma$-rays, flux variations
as short as minutes have been observed in few sources \citep{2018ApJ...854L..26S,2019arXiv190202291M,
2013ApJ...762...92A,2011ApJ...730L...8A,2007ApJ...669..862A,2007ApJ...664L..71A}.
One of the models to explain the observed flux variations in blazars is the 
shock-in-jet model, which was first proposed by \cite{1985ApJ...298..114M} and recently developed further by\citep{2010ApJ...711..445B}. Other models that
explain blazar variability include jet-star interaction \citep{2012ApJ...749..119B} 
and the magnetic reconnection models \citep{2013MNRAS.431..355G}.

Blazars have been extensively studied for flux variations at multiple wavelengths, however,
the exact mechanisms that cause flux variability are not fully understood yet. Therefore,
it is needed to enhance our understanding on the flux variability characteristics of blazars. One of 
the bands of the electromagnetic spectrum where flux variability is less characterised is the 
$\gamma$-ray regime, which is attributable to the paucity of flux variability measurements
over a high number of sources. But this band needs to be explored since this is the region where
the peak of the high-energy hump of the broad-band SED of blazars lie.
Blazars have been studied for their $\gamma$-ray variability since the launch of
the {\it Fermi} Gamma-ray Space Telescope in the year 2008. However, most of the time, individual sources
were analysed for variability, which, in addition to $\gamma$-rays utilises
data from other wavelengths\citep{2009ApJ...697L..81B,2012ApJ...749..191C,2015ApJ...803...15P,
2019MNRAS.486.1781R}. There are a limited number of studies in the literature that focus on the $\gamma$-ray flux variability characteristics of  a large sample of blazars. The first study focusing on the
$\gamma$-ray flux variability of blazars is by \cite{2010ApJ...722..520A} who 
analyse 11 months of data from the {\it Fermi} Large Area Telescope (LAT) for a total of 106 objects. 
Similarly, the $\gamma$-ray flux variability of high redshift
($z$ $>$ 3) blazars has recently been investigated by
\cite{2018ApJ...853..159L}. Quasi-periodic oscillation on year-like
time scales have also been
reported from the analysis of the long term $\gamma$-ray light curves
of blazars \citep{2015ApJ...813L..41A,2017ApJ...835..260Z,2019MNRAS.484.5785G,2019MNRAS.487.3990B}. However, a careful re-analysis of the same data set for a few objects for which quasi-periodicities were reported
did not yield any solid evidence as to the existence of year-long 
periodicities in the $\gamma$-ray light curves \citep{2019MNRAS.482.1270C,2017A&A...601A..30C}. 

The number of blazars that are known to be emitters
of $\gamma$-rays has drastically increased since the first study; additionally, $\gamma$-ray data spanning
more than ten years is now available. The availability of a homogeneous data set on a large sample
of blazars enables one to undertake a wide range of analysis in order to 
characterise $\gamma$-ray
variability of blazars.  Therefore, the main motivation of our present study is to characterise
the long term (on month-like time scales) $\gamma$-ray variability nature 
of blazars,
which includes characterising the flux variability amplitude and flux variability 
time scale that could put constraints on blazar emission models, in principle.
In addition to characterising variability, we also looked for a correlation in 
variability with
other physical properties of the sources such as the mass of 
the black hole (M$_{BH}$) and Doppler factor ($\delta$).
A description of the sample and the data used in this work is given in Section 2. The details of the
data reduction is given in Section 3, while the analysis of the data is presented in Section 4. The results
are summarised in the final section.

\begin{figure}[h]
\vbox{
      \hspace*{-0.1cm}\includegraphics[scale=0.55]{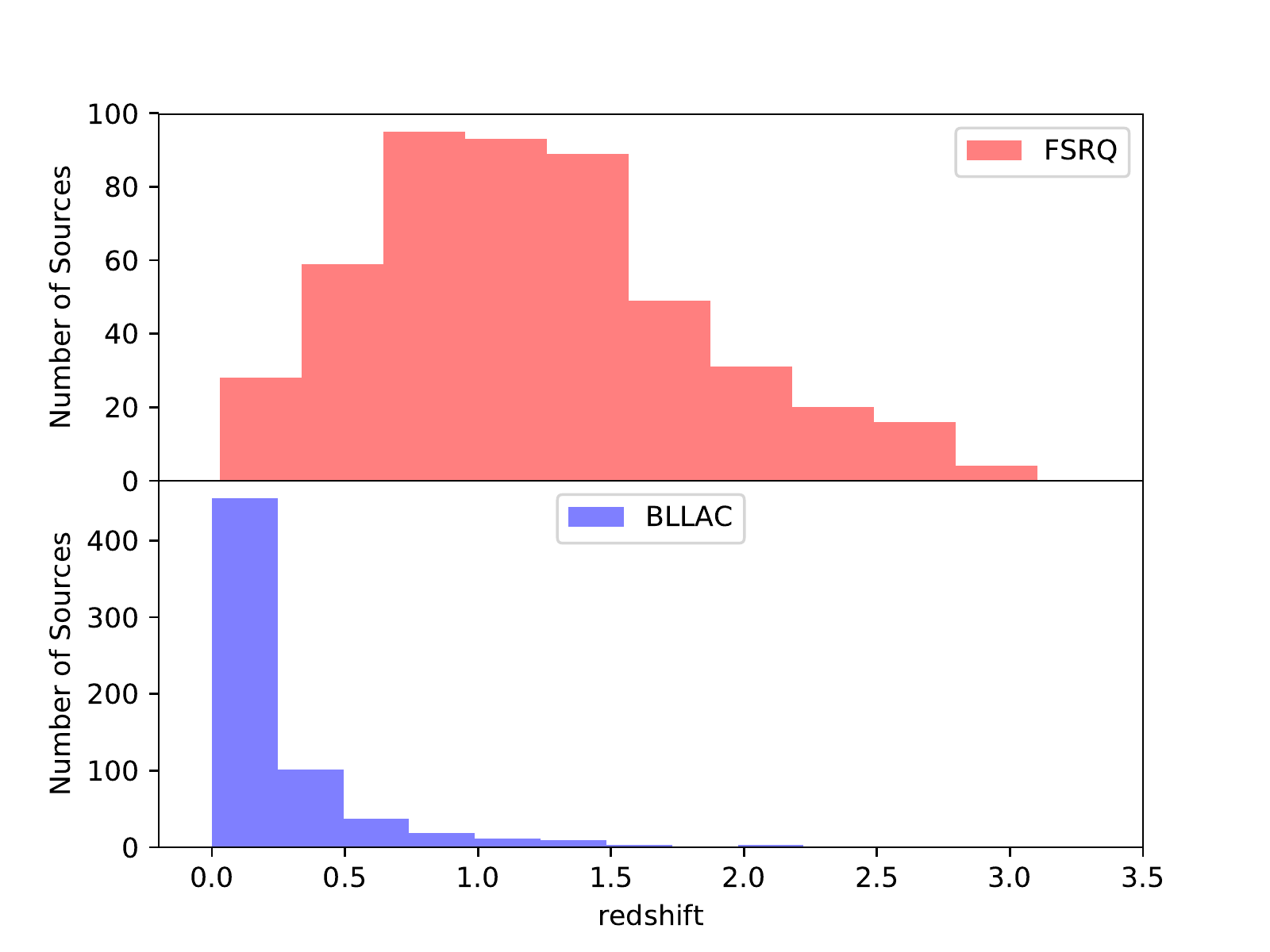}
      \hspace*{-0.1cm}\includegraphics[scale=0.55]{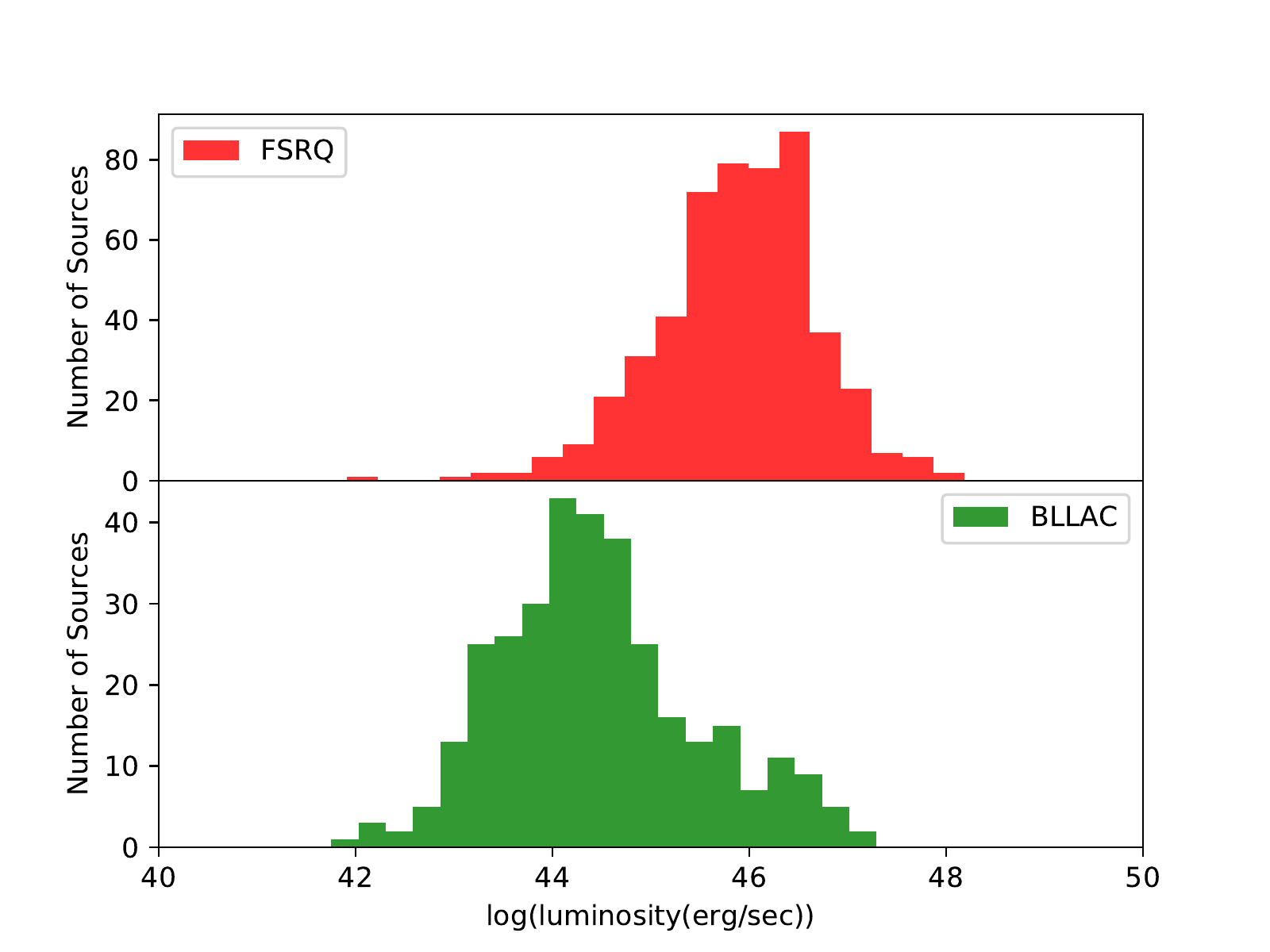}
      \hspace*{-0.1cm}\includegraphics[scale=0.55]{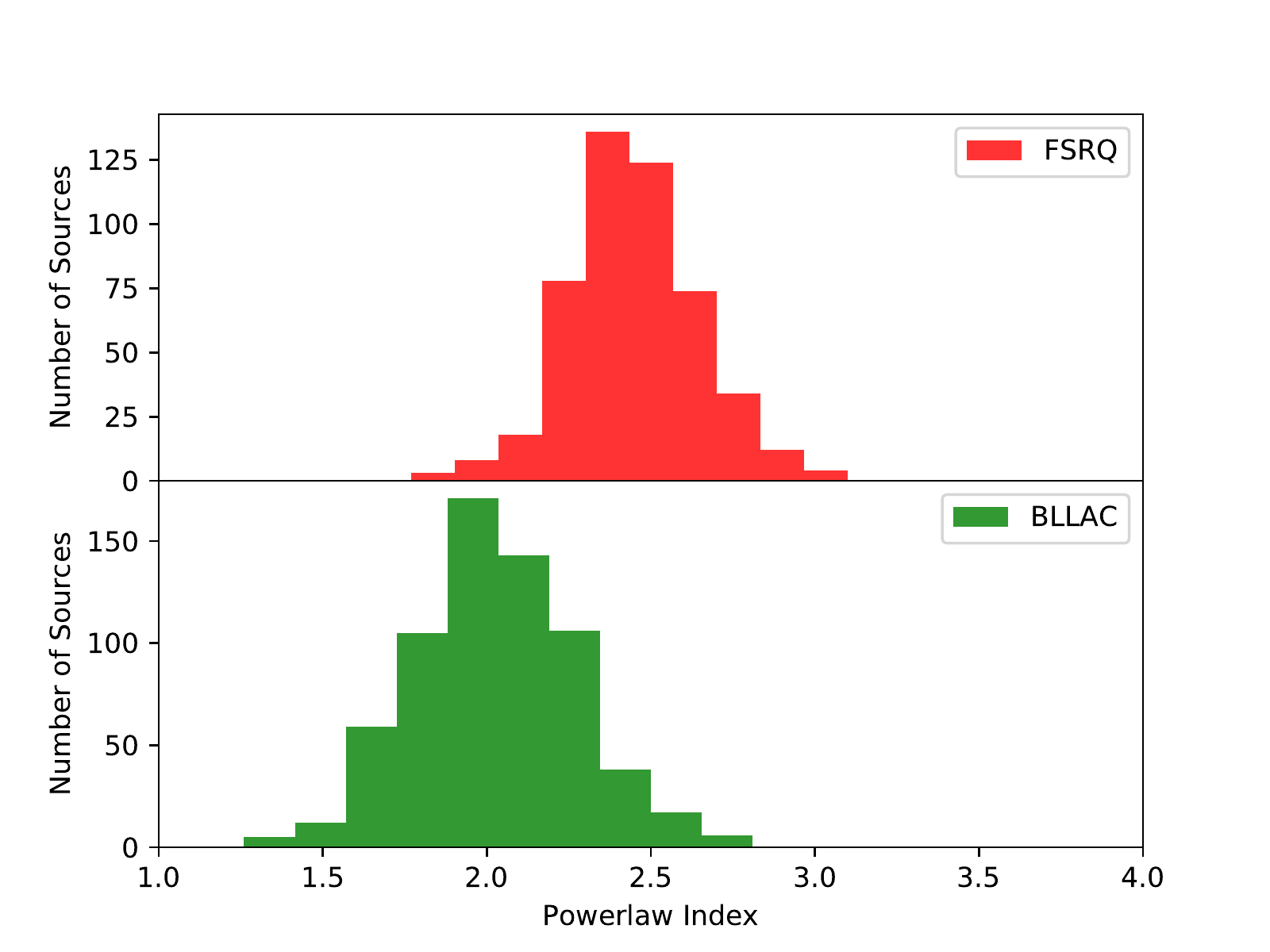}
     }
\caption{Distribution of redshifts (top panel), $\gamma$-ray luminosities in the 100 MeV-300 GeV band
 (middle panel) and $\gamma$-ray photon indices (bottom pane) for FSRQs and BL Lacs analyses 
in this work for variability}
\label{fig:fig-1}
\end{figure}

\begin{figure*}[h]
\hspace*{-2.0cm}\includegraphics[scale=0.45]{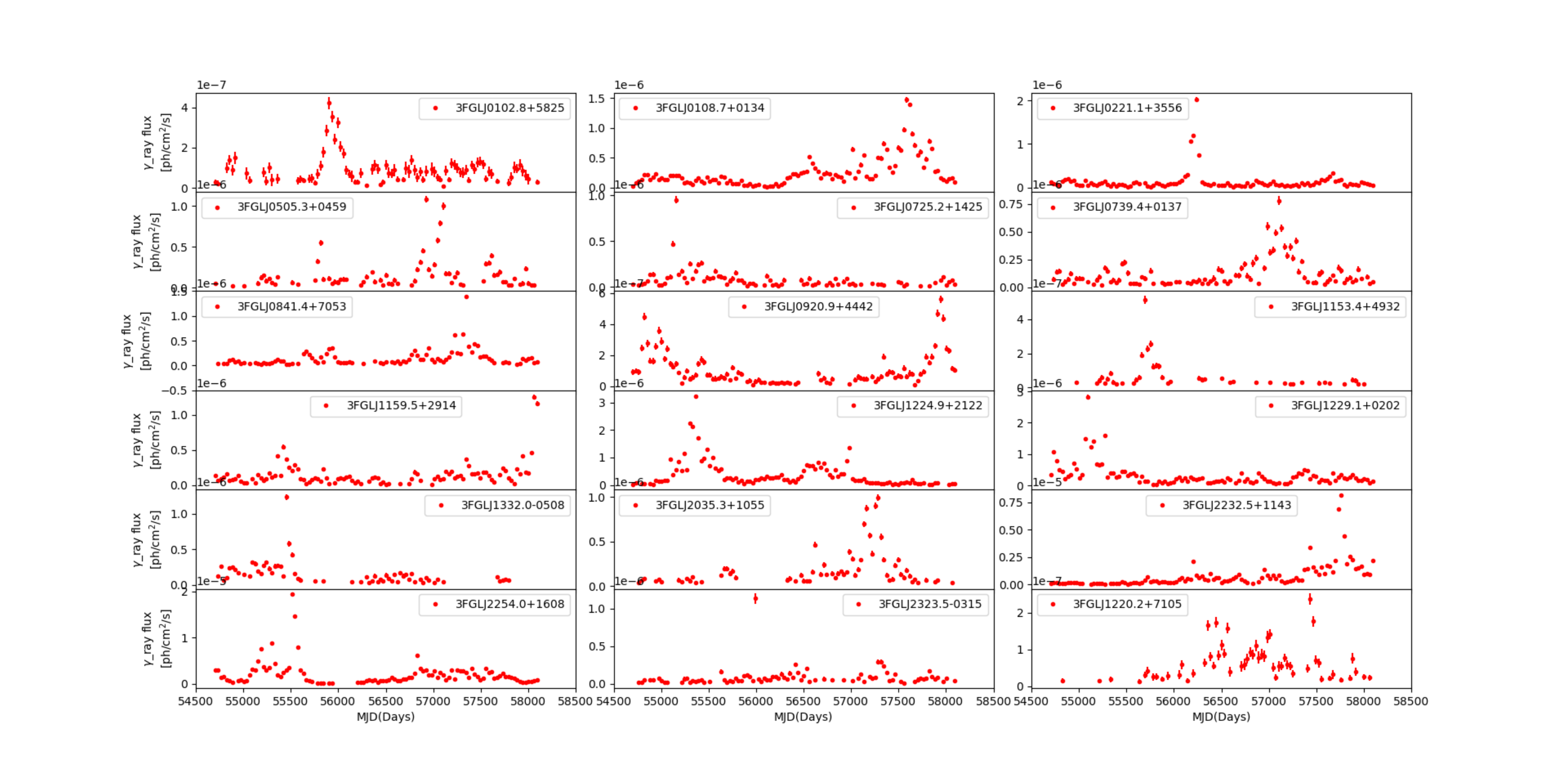}
\caption{Example light curves for variable FSRQs. The light curves generated on 
monthly time bins have their integrated fluxes measured between 100 MeV-300 GeV. The points are the flux values in the monthly bins with TS $>$ 9 
(approximately 3 $\sigma$) and the error bars are their 
1 $\sigma$ values. The names of the sources are given in each panel.}
\label{fig:fig-2}
\end{figure*}

\section{Sample and data}
The sample for our study was taken from the third catalogue of
AGN detected by {\it Fermi}-LAT (3LAC; \citealt{2015ApJ...810...14A}).
For this work we selected a total of 1120 sources detected between 100 MeV and
300 GeV with test statistic (TS) $>$ 25. The TS is a measure of source
detection significance and is defined as TS = 2$\Delta$log(likelihood)
between models with and without the source (Mattox et al. 1996).
Of these 1120 sources, 639 are BL Lacs and 481 are FSRQs. About 50\% of the
BL Lacs in our sample have no measured redshift. Excluding those
objects, the BL Lacs in our sample have redshifts between  
0.03 and 1.72, while the FSRQs have redshifts between 0.16 and 3.10.
The distribution of the redshifts of our
sample is shown in Fig. \ref{fig:fig-1}. 
By further dividing the sources in the sample that were selected
for this study and based on the position of synchrotron peak frequency in their
broad-band SED, we have 599 LSPs, 232 ISPs and 289 HSPs. 
Also shown in 
Fig. \ref{fig:fig-1} 
are the distributions of the $\gamma$-ray luminosity in the 1 $-$ 100 GeV range and the $\gamma$-ray
photon index. The $\gamma$-ray luminosities and the photon indices were taken 
from the 3LAC catalogue\footnote{https://www.ssdc.asi.it/fermi3fgl/}. FSRQs are 
highly luminous and have steeper photon indices in the $\gamma$-ray band 
relative to BL Lacs, which is similar to what is known  based on the analysis 
of three months of data from {\it Fermi} \citep{2009MNRAS.396L.105G}

\section{\bf{Data and Reduction}} \label{sec:data}

The LAT is the primary instrument on the {\it Fermi} $\gamma$-ray 
Space Telescope, which is designed to measure the energies, directions, and 
arrival times of $\gamma$-rays incident over a wide field of view and it also 
rejects cosmic-rays from the background. The LAT covers the energy range from below 
20 MeV to more than 300 GeV. The LAT has a very wide field of view 
\citep{2009ApJ...697.1071A}, very good angular resolution and good sensitivity 
over a large field of view of 2.4 steradian.  It's effective area at normal 
incidence is 9500 $cm^{2}$. The LAT is a pair-conversion $\gamma$-ray 
telescope. 
The primary observing mode of {\it Fermi} is 'scanning' mode. In this mode it covers 
the full sky in $\sim$ 3 hrs.

In this work we collected the data from 2008 August 11 to 2017 December 31 for
more than nine years within the energy range from 100 MeV to 300 GeV. We analysed 
the data using the {\it Fermi} Science Tool version v10r0p5 with the appropriate 
selections for the scientific analysis of PASS8 
data\footnote{ http://fermi.gsfc.nasa.gov/ssc/data/analysis/documentation/}. To 
analyse the data we used the publicly available python tool 
{\it fermipy} \citep{2017arXiv170709551W}. We considered the data set within the 
$15^{\circ}$ region of interest. In order to avoid background contamination, 
earth limb were excluded from the analysis (corresponding to the zenith angle cut 
of more than $90^{\circ}$). The analysis was done by using the maximum 
likelihood method (gtlike) with the instrument response function 
'P8R2$\_$SOURCE$\_$V6' , the Galactic diffuse model 'gll$\_$iem$\_$v06.fit' and the isotropic background model 'iso$\_$P8R2$\_$SOURCE$\_$V6$\_$v06.txt'. The good 
time intervals (GTIs) were created using the criteria 
'(DATA$\_$QUAL $>$ 0)\&\&(LAT$\_$CONFIG==1)'. We generated 1 month binned 
light curves for all of the sources in our sample.

\section{\bf{Analysis}} \label{sec:analysis}
\subsection{\bf{Monthly Binned Light curves}}
\begin{figure*}[h]
\hspace*{-2.0cm}\includegraphics[scale=0.45]{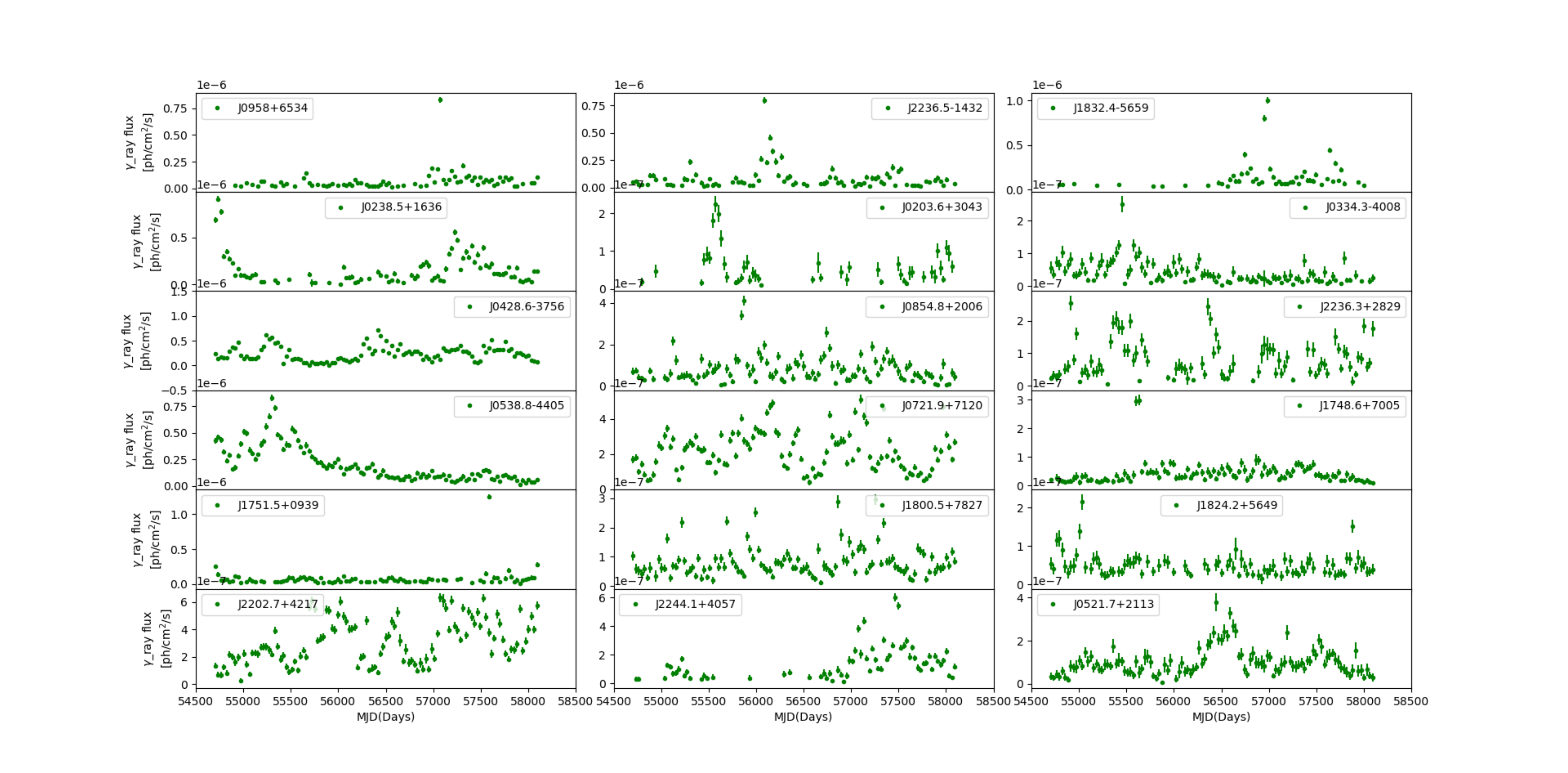}
\caption{Example monthly binned light curves (TS $>$ 9) along with their 
1 $\sigma$ errors for BL Lacs. The names of the sources
are given in each panel. Each point in the light curves refers to flux measured in the
100 MeV-300 GeV band}
\label{fig:fig-3}
\end{figure*}

The $\gamma$-ray light curves of our sample of sources were generated 
as per the details found in Section 3 for a period of about nine years from 2008
August 11 to 2017 December 31. The light curves were generated with a time binning of one month which results in 114 bins for each light curve. For each interval we calculated 
the flux and test statistic (TS) values for every source. The TS values were 
calculated using the maximum likelihood function gtlike. 
We considered a source to 
be detected at any epoch if its TS $>$ 9 (3$\sigma$ detection). At epochs
where TS $<$ 9, the source was considered undetected. In Figure \ref{fig:fig-2} 
and Figure \ref{fig:fig-3}, we show the light curves of a few FSRQs and  
BL Lacs from our sample. It is likely that many light curves do not have
flux measurements every month and missing flux points are due to the 
source's flux below our detection threshold.

\begin{figure}
\hspace*{-0.2cm}\includegraphics[scale=0.6]{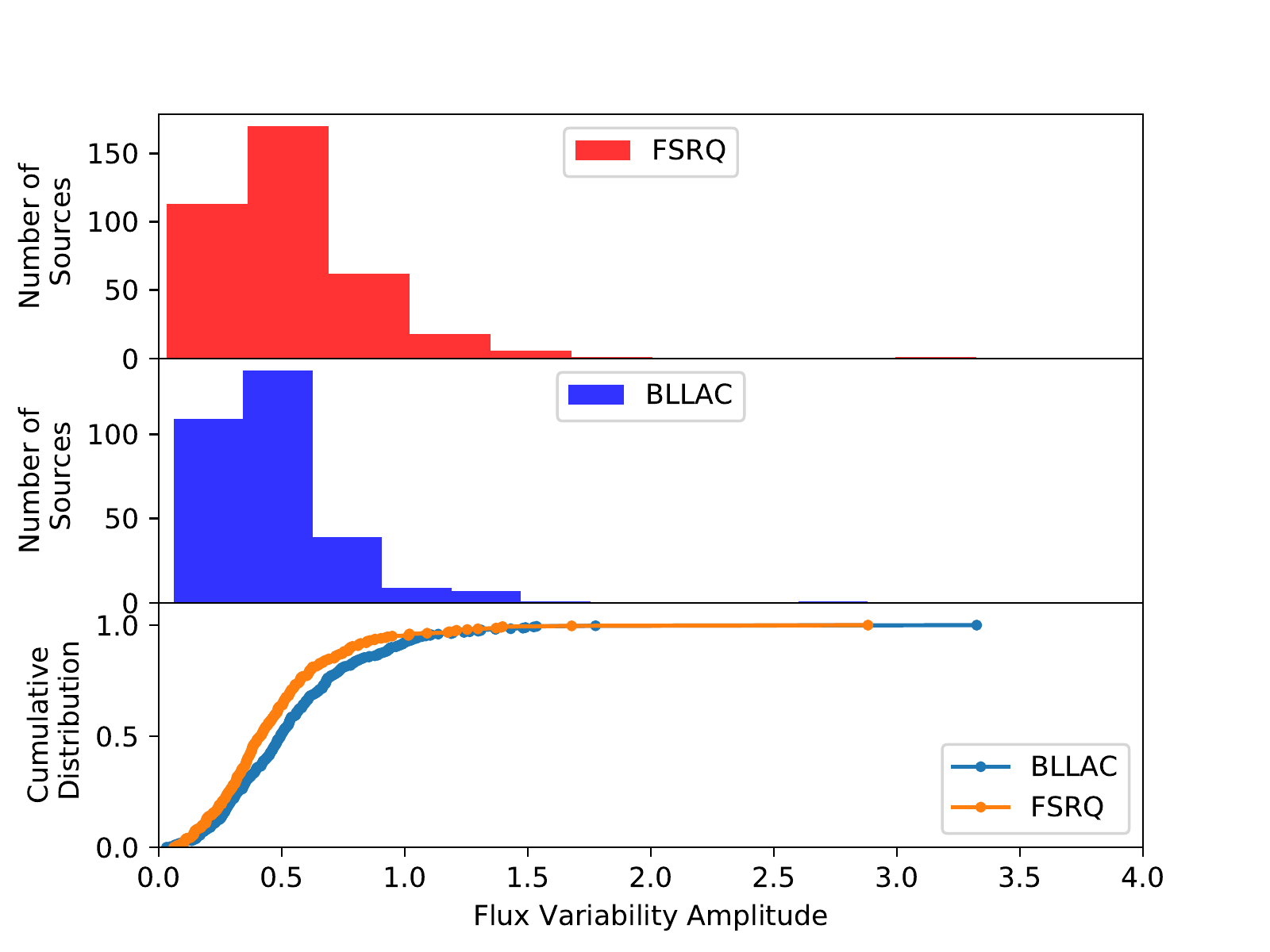}
\caption{Histogram and cumulative distribution of F$_{var}$ for variable FSRQs and 
BL Lacs studied in this work }
\label{fig:fig-4}
\end{figure}

\begin{figure}
\hspace*{-0.2cm}\includegraphics[scale=0.6]{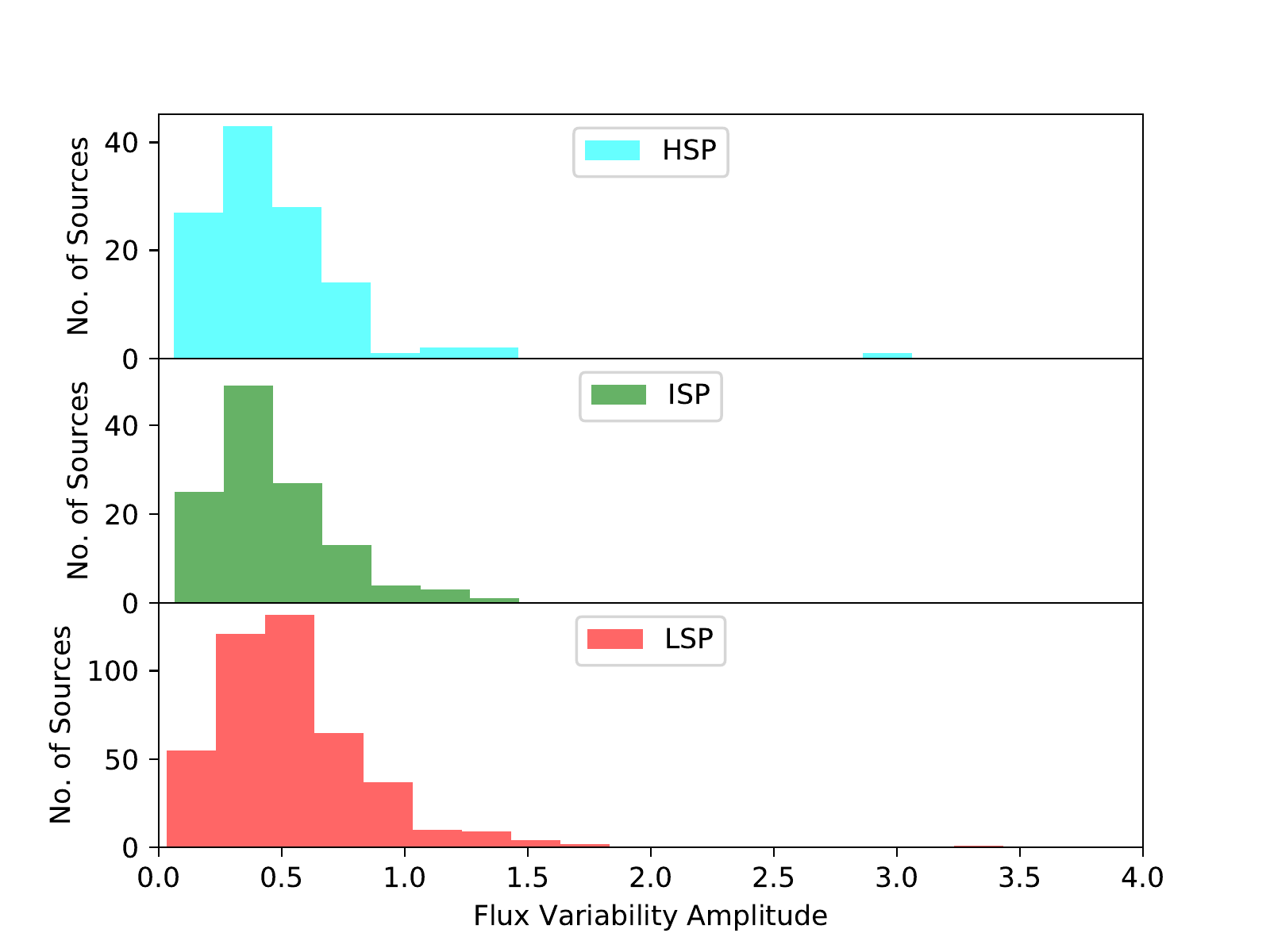}
\caption{Distribution of F$_{var}$ values for variable  LSP, ISP and HSP blazars in 
our sample}
\label{fig:fig-5}
\end{figure}

\begin{figure}
\hspace*{-0.2cm}\includegraphics[scale=0.6]{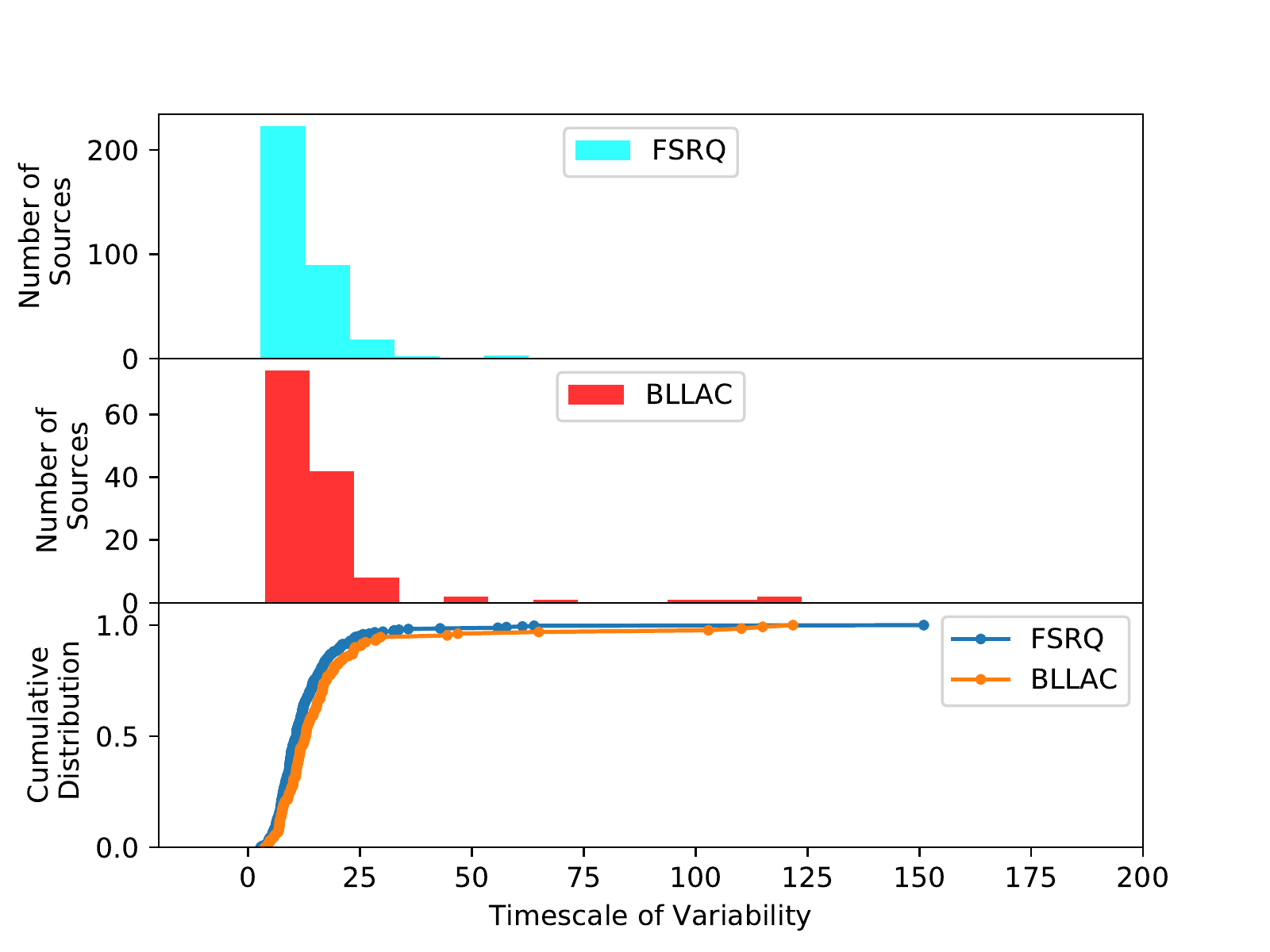}
\caption{Histogram and cumulative distribution of the time scale of variability (days) for FSRQs and BL Lacs.}
\label{fig:fig-6}
\end{figure}

\begin{figure}
\centering
\resizebox{8cm}{7cm}{\includegraphics{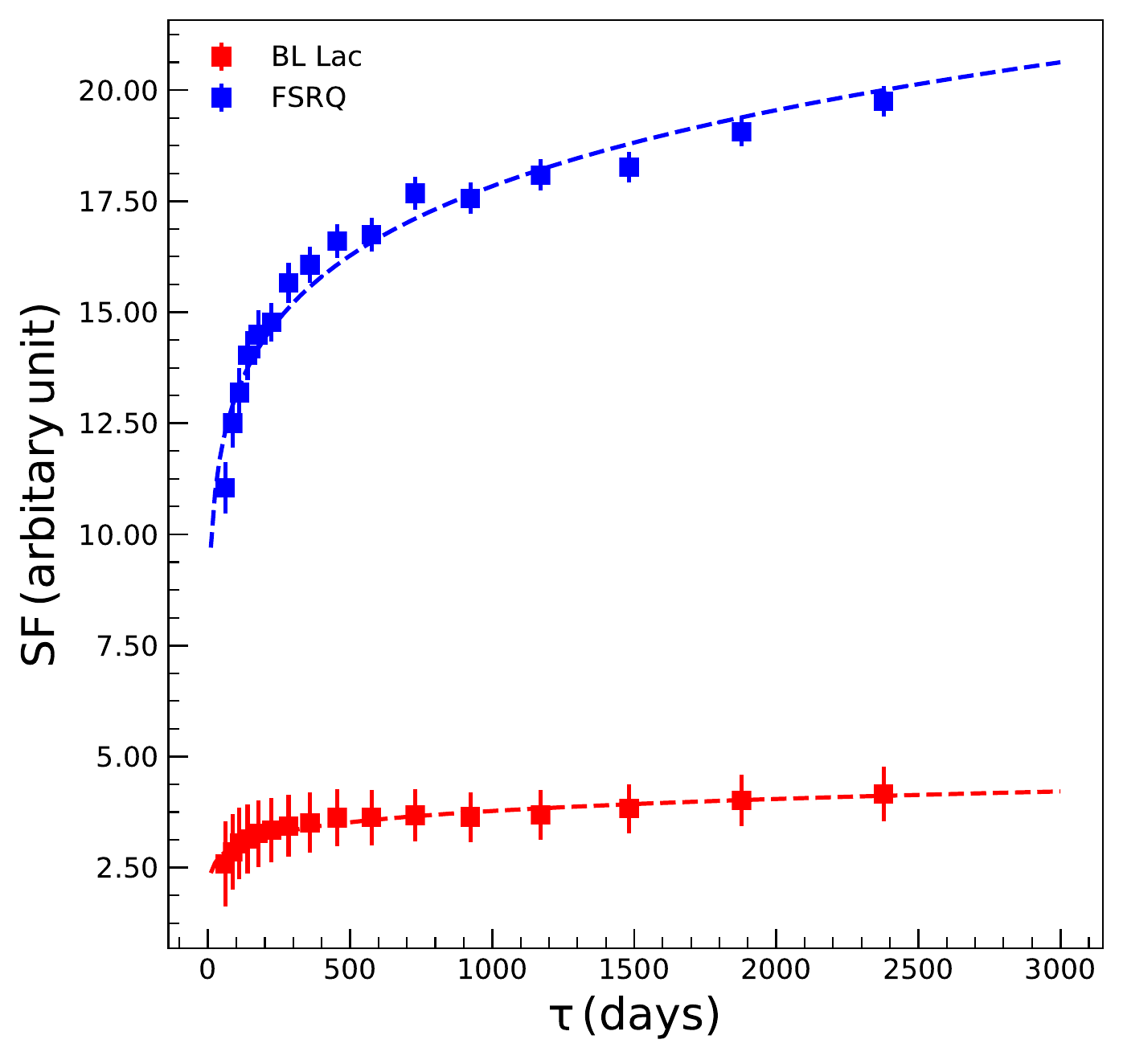}}
\caption{Structure function (SF) against observed frame time lag for BL Lacs 
(red dots) and FSRQs (blue dots). The dashed lines are the best fits to the SF using Equation 9.}
\label{fig:fig-7}.
\end{figure}

\begin{figure}
\centering
\resizebox{8cm}{7cm}{\includegraphics{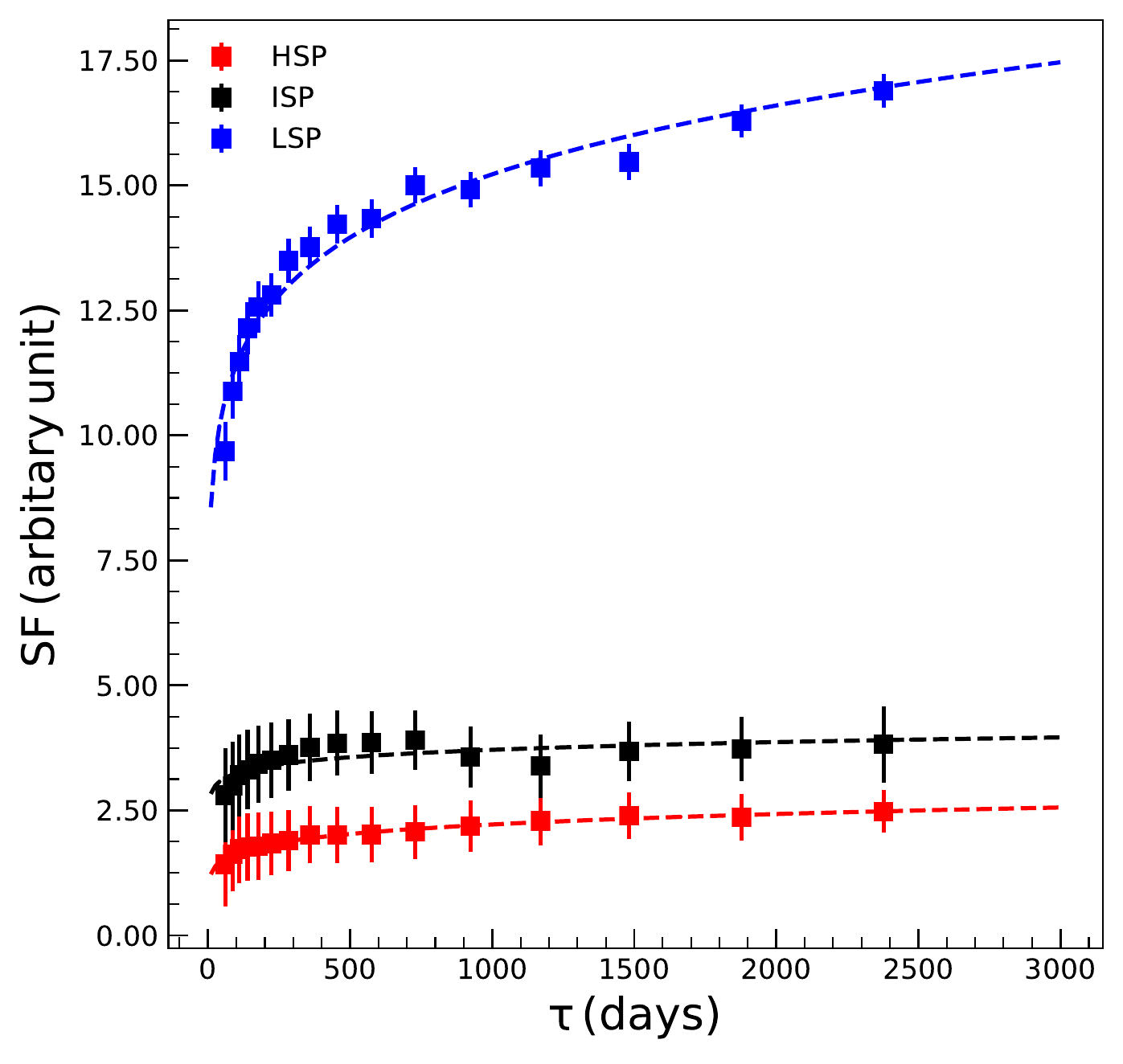}}
\caption{Structure functions for HSP (red), ISP (black) and LSP (blue) blazars. 
The dashed lines are the best fits to the SF using Equation 9.}
\label{fig:fig-8}.
\end{figure}

\begin{figure}
\hspace*{-0.2cm}\includegraphics[scale=0.45]{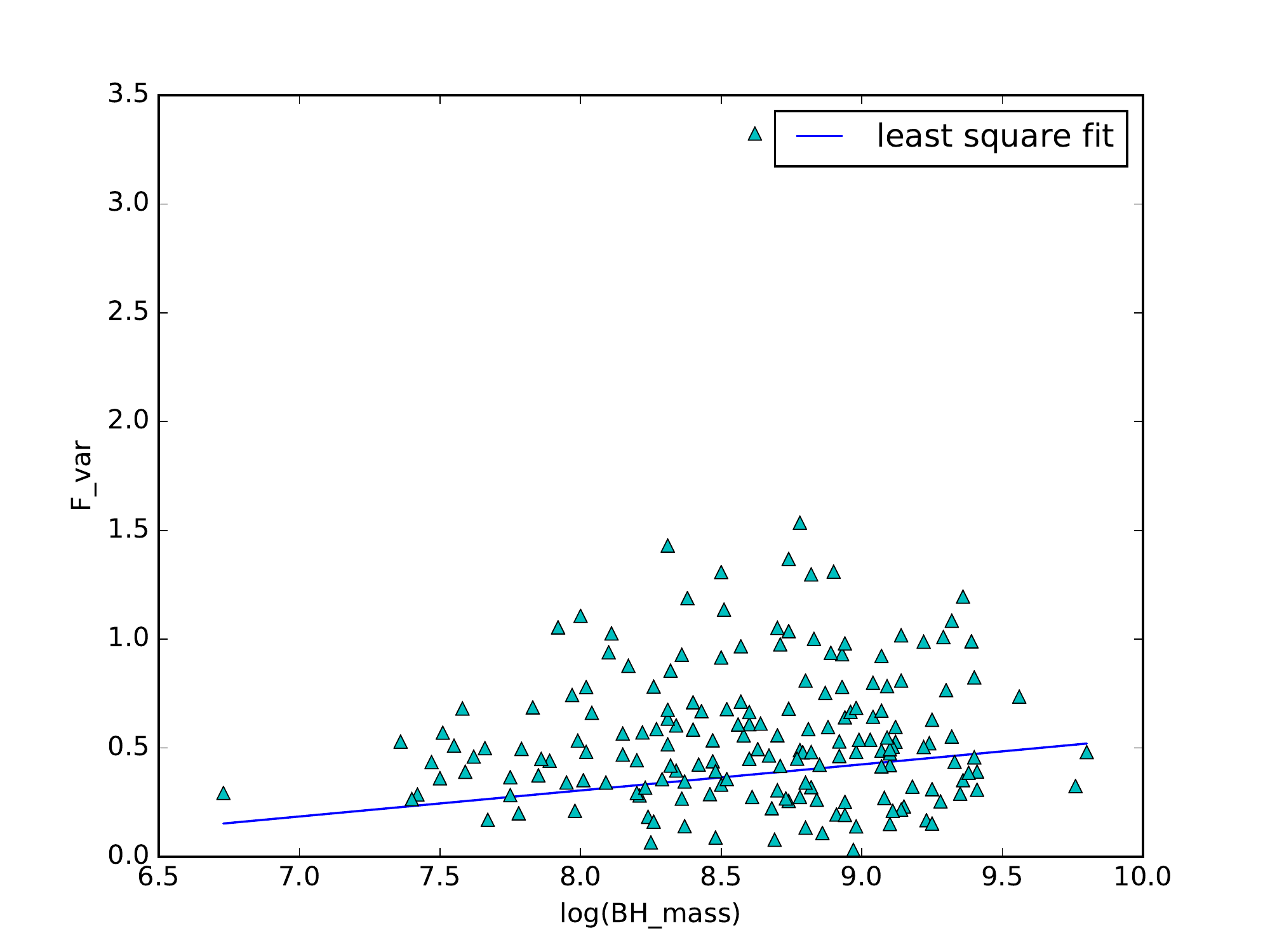}
\caption{Correlation between F$_{var}$ and M$_{BH}$ values for FSRQs. The solid line
is the unweighted linear least squares fit to the data}
\label{fig:fig-9}
\end{figure}

\begin{figure}
\hspace*{-0.2cm}\includegraphics[scale=0.6]{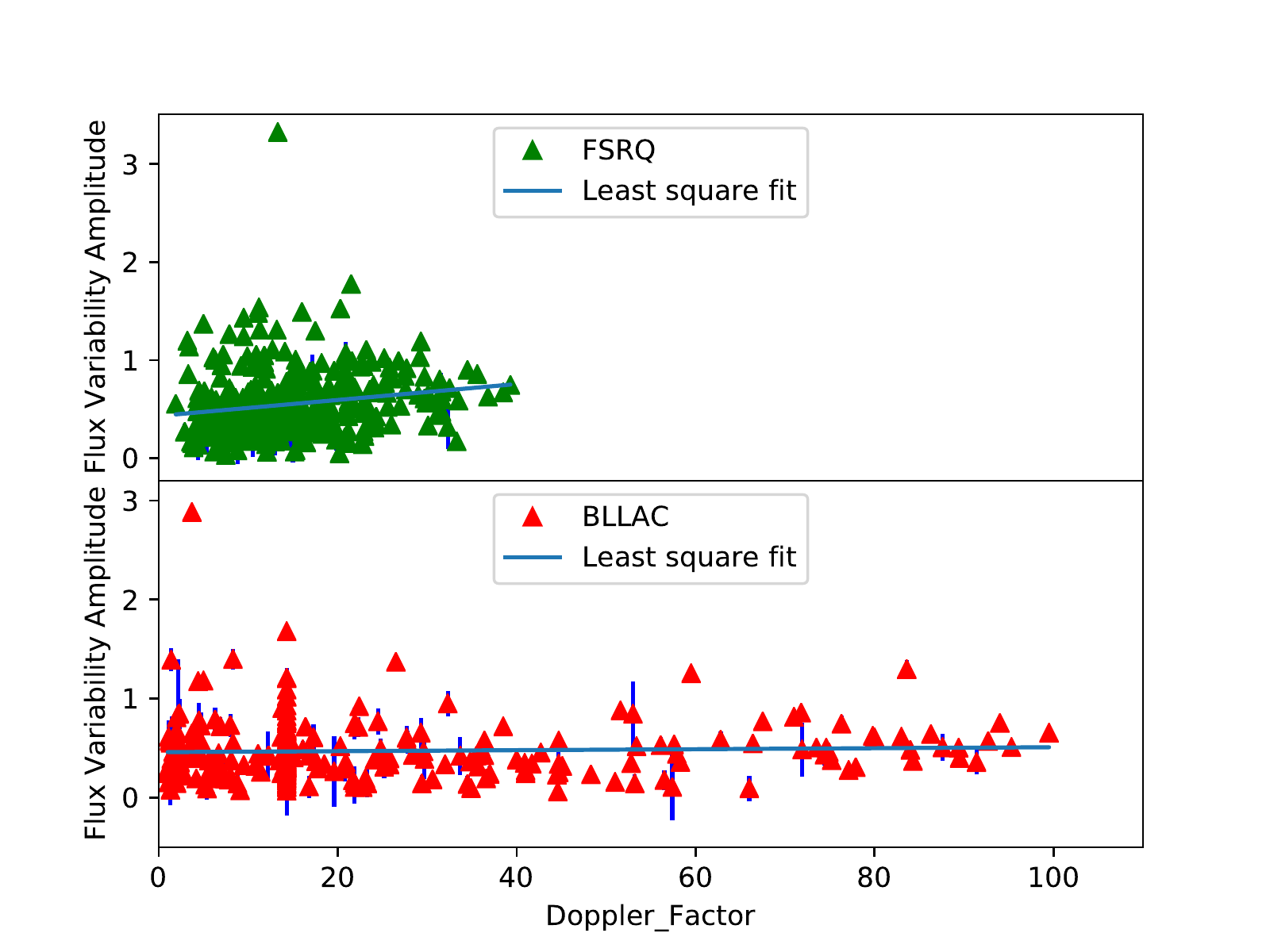}
\caption{Relation between $F_{var}$ and Doppler factor for FSRQs (top panel) and BL Lacs (bottom panel).
Unweighted linear least squares fit to the data are shown as solid lines.}
\label{fig:fig-10}
\end{figure}

\begin{figure}
\hspace*{-0.2cm}\includegraphics[scale=0.6]{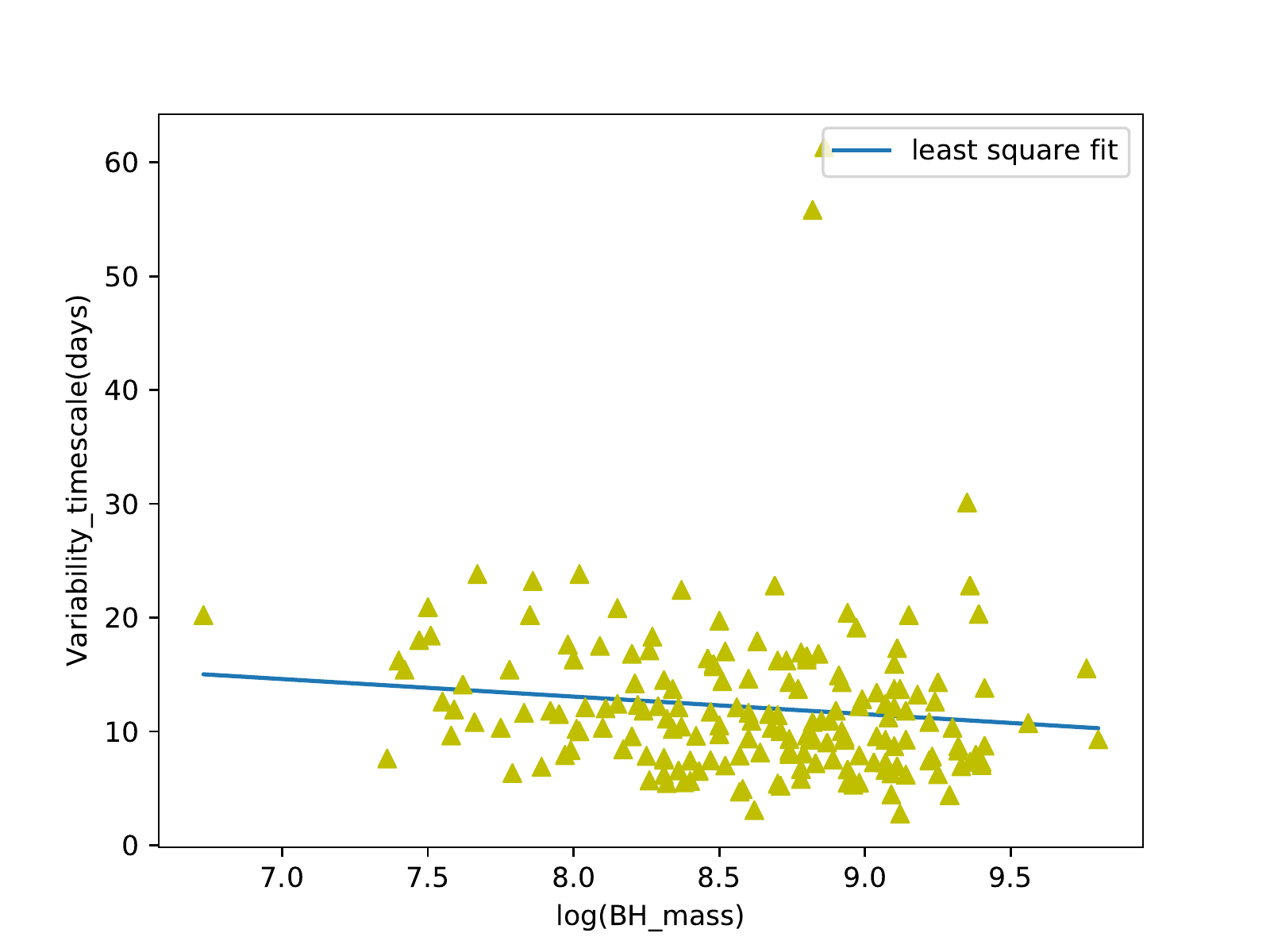}
\caption{Correlation between time scale of variability and M$_{BH}$ for blazars. The solid
line is the unweighted linear least squares fit to the data points.}
\label{fig:fig-11}
\end{figure}

\begin{figure}
\hspace*{-0.2cm}\includegraphics[scale=0.6]{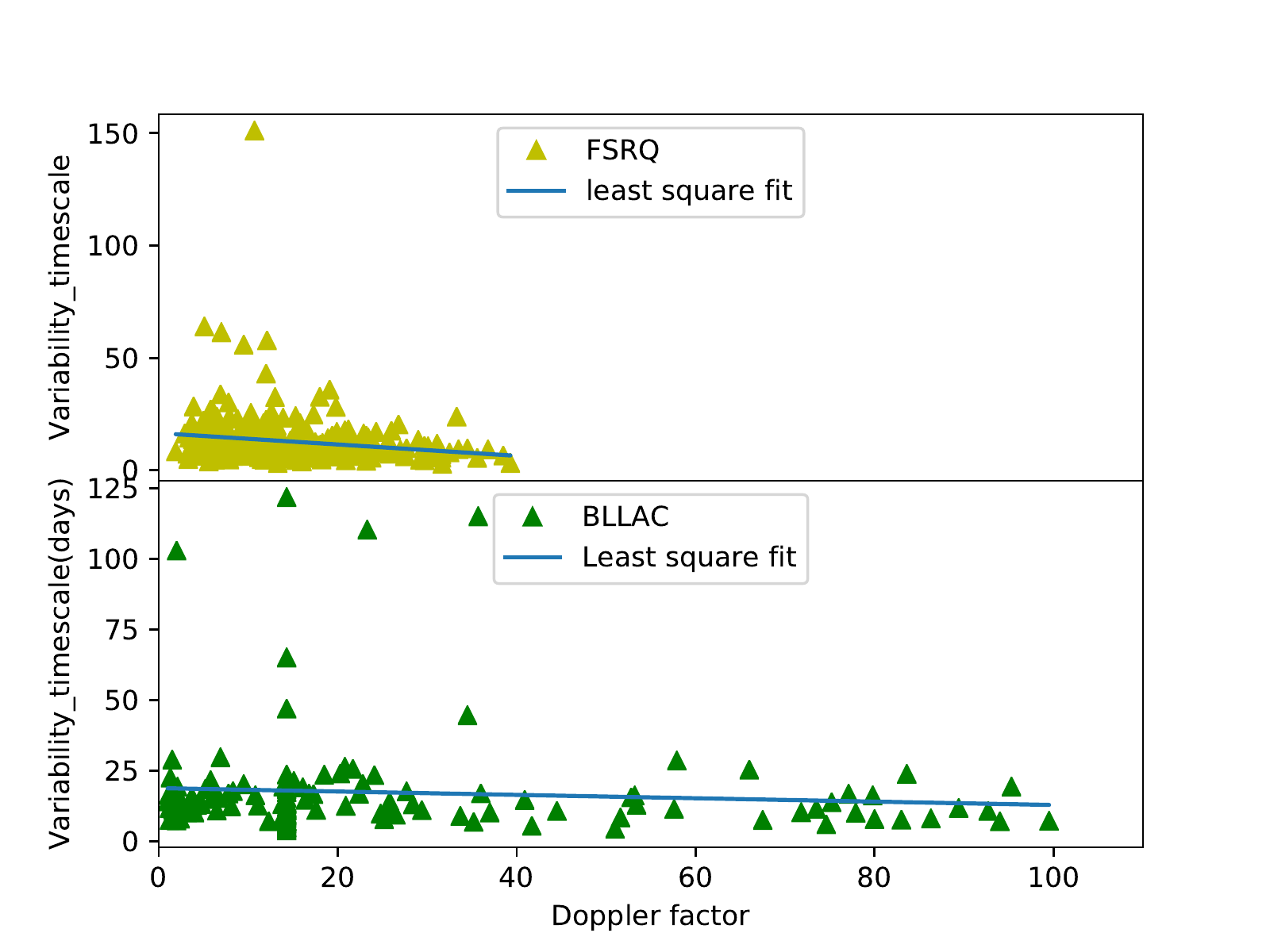}
\caption{Correlation between time scale of variability and Doppler factor for FSRQs(top panel) and
BL Lacs (bottom panel). Linear least squares fit to the data are shown as solid lines.}
\label{fig:fig-12}
\end{figure}

\subsection{\bf{Flux Variability Amplitude}}
To quantify flux variability, we used the fractional root mean square
variability amplitude ($F_{var}$;\citealt{2003MNRAS.345.1271V}). This is
defined as

\begin{equation}
F_{var} = \sqrt{\frac{S^{2} - \bar{\sigma^{2}_{err}}}{\bar{x}^{2}}}
\end{equation}

where $S^{2}$ is the sample variance and  $\bar{\sigma^{2}_{err}}$ is mean 
square error. They are given as
\begin{equation}
S^{2} = \frac{1}{N-1}\sum_{i=1}^{N}(x_{i} - \bar{x})^{2}
\end{equation}
and
\begin{equation}
\bar{\sigma^{2}_{err}} = \frac{1}{N}\sum_{i=1}^{N}{\sigma^{2}_{err,i}}
\end{equation}

Here, $\sigma_i$ is the statistical uncertainty, to which we added the systematic uncertainty
$\sigma_{syst}$ = 0.03 <x$_i$> in quadrature  \citep{2009ApJ...700..597A} to get the
total error $\sigma_{err}$ defined as

\begin{equation}
\sigma_{err}^{2} = \sigma_{i}^{2} + \sigma_{sys}^{2}
\end{equation}

The uncertainty in $F_{var}$ is defined as \citep{2017MNRAS.466.3309R}
\begin{equation}
err(F_{var}) = \sqrt{\Bigg(\sqrt{\frac{1}{2N}}\frac{\bar{\sigma_{err}^{2}}}{\bar{x}^{2}F_{var}}\Bigg)^{2} + \Bigg(\sqrt{\frac{\sigma_{err}^{2}}{N}}\frac{1}{\bar{x}}\Bigg)^{2}}
\end{equation}

In Fig. 4, the distribution  and cumulative distribution of $F_{var}$ for FSRQs and BL Lacs are shown. We found mean $F_{var}$ values
of 0.47 $\pm$ 0.29 and 0.55 $\pm$ 0.33 for BL Lacs and FSRQs, respectively. A two sample Kolmogorov Smirnov (KS) test shows that
the two distributions are indeed different at the 95\% level with statistics of 0.15 and a $p$ value of 0.001. 
We also sub-divided the sample into different spectral energy distribution 
classes based on the peak frequency of the low energy
synchrotron component in their broad-band SED. 
The mean $F_{var}$ values for the different sub-classes are
0.54 $\pm$ 0.33 for LSPs, 0.45 $\pm$ 0.25 for ISPs, and 0.47 $\pm$ 0.33 for 
HSPs. The distribution of $F_{var}$ values for the
different sub-classes are shown in Fig. 5. \cite{2011ApJ...743..171A} also
find a similar trend of flux variations in the $\gamma$-ray band for 
different classes of blazars. By only Considering BL Lacs, \cite{2011ApJ...743..171A},
find that variability decreases from LSP to ISP and HSP.

\subsection{Duty cycle of variability}
We calculated the duty cycle (DC) of variability, including only those sources
that have a redshift measurement, in order to determine the fraction of 
time a particular class of sources shows flux variations. The DC was estimated
following \cite{1999A&AS..135..477R} and is given as
\begin{equation}
DC = 100 \frac{\Sigma_{i=1}^N{Q_i (1/\Delta t_i)}}{\Sigma_{i=1}^{N} (1/\Delta t_i)}
\end{equation}
where $\Delta t_i = \Delta t_i (1 + z)^{-1}$ is the time in the rest frame
of the source, $N_i$ = 1 if a particular source is variable, or else $N_i$ = 0. For FSRQs, we find a DC of 66\%, while for BL Lacs, we find 
a DC of 36\%. For the sub-classes of blazars we find DCs of 65\%, 43\%, and
36\% for LSP, ISP, and HSP blazars, respectively. Thus, LSP sources show a
larger DC of $\gamma$-ray variability on month-like time scales related 
to the other classes of blazars.

\subsection{\bf{Variability timescale}}

The variability time scale ($\tau$) is a very important parameter that can be 
deduced from the light curves, which in turn can provide constraints on the 
physical processes that cause $\gamma$-ray flux variations. Since we analyses monthly binned light curves in this work, we were able to 
probe time scales of the order of months. We calculated $\tau$ 
of $\gamma$-ray 
flux variability for the sources in our sample that showed 
$\gamma$-ray flux variability following \cite{2013ApJ...773..147J}
\begin{equation}
\tau \equiv \bigtriangleup t/ln(S_{2}/S_{1})
\end{equation}

Here $S_{2}$ and $S_{1}$ are flux values at a time of $t_{2}$ and $t_{1},$ 
respectively, and $\Delta$t =  $\mid$$t_{2}$-$t_{1}$$\mid$. In order to estimate $\tau,$ we considered all possible pairs of flux values 
that satisfy the conditions (i) $S_{2}$ $>$ $S_{1}$ and 
(ii) $S_{2} - S_{1} > 3(\sigma_{S_{1}} + \sigma_{S_{2}}$)/2, where 
$\sigma_{S_{2}}$ and $\sigma_{S_{1}}$ are the uncertainties corresponding to
the flux measurements $S_1$ and $S_2,$ respectively. Among all of the calculated values of
$\tau$ for a particular source, we considered the minimum $\tau$ value
as the timescale of variability of the source with the $\gamma$-ray flux
changing by a factor  greater than 2. The histogram and cumulative
distribution of $\tau$ for FSRQs and BL Lacs are 
shown in Fig. \ref{fig:fig-6}.\\


\subsection{Ensemble structure function}
The variability of AGN can also be described by the structure function (SF), which 
shows the dependency of variability as a function of time lag 
\citep{1985ApJ...296...46S}. The SF can be calculated for individual AGN that have a light curve with multiple epochs of observations, which takes the magnitude difference
for each pair of time lags in a light curve. It can also be calculated for a 
group of AGN, known as the ensemble structure function, allowing us to obtain the 
mean variability behaviour of the population that is similar to what has been obtained 
from the flux variability amplitude. We studied the mean variability of different 
classes of AGN by using the ensemble structure function following 
\cite{1996ApJ...463..466D} 
\begin{equation}
\mathrm{SF} = \sqrt{\frac{\pi}{2} <|\Delta m|>^2 - <\sigma{^2}_n>},
\end{equation}
where $|\Delta m|=m_i - m_j$, is the magnitude difference between any two epochs 
($i$, $j$) that are separated by time $\Delta \tau = t_i - t_j$. 
$\sigma{^2}_n=\sigma{^2}_i+\sigma{^2}_j$, which is the square of the uncertainty of 
the magnitude differences. We note that the majority of our sources do not have 
redshift measurements in the literature, thus, the SF was calculated in the 
observed frame. In Figure \ref{fig:fig-7}, we plotted the SF against the observed frame time lag for BL Lacs (red) and FSRQs(blue). The error bar in the SF was calculated via error propagation following \cite{2004ApJ...601..692V}. 
Figure \ref{fig:fig-7} clearly shows that FSRQs are more variable than BL Lacs, which is consistent with the result obtained by $F_{var}$ analysis. The SF increases 
gradually from time lags ranging from one to $\sim$400 days and becomes flatter at higher time lags. Such a 
trend has been noted previously by various authors 
\citep{2004ApJ...601..692V,2011A&A...527A..15W,2016ApJ...826..118K}. To characterise the structure function, we fitted the following simple power-law model:

\begin{equation}
\mathrm{SF} = S_0 \times \left(\frac{\Delta \tau}{\tau_0}\right)^{\gamma}.
\label{eq:SF_PL}
\end{equation}
By adopting $\tau_0$=4 years in the observed frame \citep{2016ApJ...826..118K}
we estimated $S_0$ and $\gamma$. The fitting results are given in Table \ref
{Table:PL}. The higher value of $S_0$ in FSRQs than BL Lacs suggests the former 
has higher variability than the latter. This is also confirmed from the 
higher flux variability of the FSRQs compared to BL Lacs. In Figure 
\ref{fig:fig-8}, we show the SFs of HSP, ISP, and LSP. We find that 
LSPs have stronger variability followed by ISP and HSP blazars. This is 
also in agreement with that was obtained from the $F_{var}$ analysis. 

Based on the analysis of 106 $\gamma$-ray light curves using 11 months of 
data from {\it Fermi}, \cite{2010ApJ...716...30A} find FSRQs to show a higher
amplitude of $\gamma$-ray variability than other AGN classes. Similarly, from an analysis
of the sources in the  second LAT AGN catalogue, \cite{2011ApJ...743..171A} find FSRQs to 
have more flux variability than BL Lacs. According to \cite{2011ApJ...743..171A}, the 
higher variability seen in FSRQs relative to BL Lacs could be attributed to the
location of the high-energy peak (in the broad-band SED of blazars) with respect
to the {\it Fermi} band. In the {\it Fermi} band, FSRQs are observed at energies
greater than the inverse Compton peak in the SED; the observed emission is therefore
produced by high-energy electrons with shorter cooling time scales and thereby shows
more variations. Alternatively, in the {\it Fermi} band, BL Lacs are observed at 
frequencies much lower than the inverse Compton peak, the low-energy electrons have
longer cooling time scales, and therefore show low variations. The results obtained in 
this work on a large sample of blazars having data spanning about nine years is in agreement with the
earlier results that were obtained on a smaller sample of blazars with less time coverage
\citep{2011ApJ...743..171A,2010ApJ...716...30A}.

\begin{table}
\caption{Results of model fits to the structure function using power-law model.}
\begin{center}
\resizebox{0.9\linewidth}{!}{%
\begin{tabular}{ l r l}\hline \hline
Object class    &   $S_0$ ($10^{-8} \, \mathrm{ph\ cm{^2}\ s^{-1}}$)  &  $\gamma$  \\ \hline
    BL Lac      & $3.92  \pm 0.04$~~~~~   & $.100 \pm 0.007 $ \\
    FSRQ        & $18.70 \pm 0.20$~~~~~   & $0.132 \pm 0.007 $ \\
    HSP         & $2.33  \pm 0.02$~~~~~   & $0.129 \pm 0.006 $  \\
    ISP         & $3.79  \pm 0.08$~~~~~   & $0.058 \pm 0.014 $  \\
    LSP         & $15.95 \pm 0.16$~~~~~   & $0.124 \pm 0.007 $  \\
\hline
\end{tabular} } 
\label{Table:PL}
\end{center}
\end{table}

\subsection{$F_{var}$, M$_{BH}$, and Doppler factor}
We searched in the literature for the availability of $M_{BH}$ values for the 
sources analyses for variability here. We could gather $M_{BH}$ values \citep{2018ApJS..235...39C} for 
a total of 184 FSRQs.  In Fig. \ref{fig:fig-9}  we show $F_{var}$ as a function 
of $M_{BH}$ for  FSRQs. There is a weak indication of larger $\gamma$-ray 
flux variations in sources with large $M_{BH}$ values. However, linear least squares fit to the 
data showed an insignificant correlation between $F_{var}$ and $M_{BH}$ with a linear
correlation coefficient of 0.07.
\cite{2001MNRAS.324..653L} carried out an analysis of the X-ray flux
variations on a composite sample of Seyfert 1 galaxies, quasars and narrow line
Seyfert 1 galaxies and found  a significant anti-correlation between X-ray
variability and $M_{BH}$. Upon the analysis of the long term optical variability
characteristics of a large sample of quasars, \cite{2012ApJ...758..104Z}  could not
find any correlation between $M_{BH}$ and variability amplitude, however, other
studies have found a correlation between quasar variability and $M_{BH}$
\citep{2007MNRAS.375..989W,2009ApJ...696.1241B}, while \cite{2009ApJ...698..895K} find a negative
correlation between M$_{BH}$ and quasar variability. \cite{2010ApJ...716L..31A}
note that the correlation between optical variability and M$_{BH}$ vanishes when the Eddington ratio is controlled.

The correlation between $F_{var}$ and $\delta$ for FSRQs and BL Lacs 
is shown in Fig. \ref{fig:fig-10}. We note that $\delta$ was also collected from \cite{2018ApJS..235...39C}. The figure is suggestive of a positive correlation
between $F_{var}$ and $\delta$. However, from the linear least squares fit to the 
data points, we find no correlation between $F_{var}$ and $\delta$ in both
FSRQs and BL Lacs. Any small changes in the jet emission in blazars would get 
Doppler boosted, leading to the large amplitude of flux variations by the observer. Even though our data sets are indicative of such a correlation, no clear trend could 
be established.

\subsection{Time scale of variability, M$_{BH},$ and Doppler factor}
Knowledge on the time scale of flux variations in blazar light curves is very
important as it can provide us important clues as to the physical processes
responsible for $\gamma$-ray flux variations in blazars. The
power spectral density (PSD) is generally used to quantify the time scale
of flux variations in blazars, however, we followed the approach given
in Eq.7 to determine the time scale of variability in the monthly
binned blazar light curves. From a homogeneous analysis of the
blazar light curves, we find that most of the sources analyses in this work
have a time scale of variability that is less than 50 days, while few
sources have time scales larger than 100 days.
From a PSD analysis of the weekly and daily binned $\gamma$-ray light curves of 13 blazars spanning
about ten years, \cite{2019arXiv190904227R}  observed two time scales of variability,
the longer time scale having a duration of the order of years and the
shorter time scale spanning of the order of days. According to them, 
the longer time scales might represent the thermal time scale of the accretion
disc, while the shorter time scales may be related to processes in the jet.
For most of the sources analyses here, the estimated time scales are
of the order of days, and such time scales could be related to emission
processes in the jet \citep{2019arXiv190904227R}.

Even though, historically, blazars are separated into FSRQs and BL Lacs based
on the width of the emission lines present in their optical spectrum,
\cite{2009MNRAS.396L.105G} postulate a physical
distinction between FSRQs and BL Lacs. The PSDs associated with EC, which
produces $\gamma$-ray emission in FSRQs, and SSC, producing $\gamma$-ray emission in BL Lacs, show different break frequencies \citep{2019arXiv190904227R}
. In such a scenario, different time scales of
variability in the $\gamma$-ray band are expected. The distribution of
$\tau$ for both FSRQs and BL Lacs are shown in Fig. \ref{fig:fig-4}. A
KS test indicates that the two distribution are marginally different,
with a statistic of 0.18 and $p$ values of 0.004. We thus noticed a difference in
the distribution of the time scales of variability  between FSRQs and BL Lacs.

The correlation between $\tau$ and M$_{BH}$ in blazars were found in the X-ray
\citep{2018ApJ...859L..21C} and optical
\citep{2009ApJ...698..895K,2010ApJ...721.1014M}. In Fig. \ref{fig:fig-11}, we
show the correlation between $\tau$ in the $\gamma$-ray band against $M_{BH}$. The linear least squares fit to the data yields a low correlation coefficient of $-$0.12. We therefore do not find a significant correlation between $\tau$ and M$_{BH}$.
We also do not find any  correlation between $\tau$ 
and $\delta$ for both FSRQs and BL Lacs (Fig. \ref{fig:fig-12}). 
Doppler boosting shortens the observed time scale by $\delta^{-1}$, and
the observed hint (though insignificant) of a  negative correlation is a consequence 
of the effect of $\delta$ on the time scale of flux variations.


\section{Summary}
In this work we generated one month binned $\gamma$-ray light curves for a 
total of 1120 blazars, comprising 481 FSRQs and 639 BL Lacs to characterise 
their $\gamma$-ray variability with the data collected from
{\it Fermi} for over approximately nine years. This is a systematic study of the $\gamma$-ray 
flux variability using a large sample of blazars.
The results of this work are summarised below
\begin{enumerate}

\item More than 50\% of the blazars studied in this work are found  to be variable. Out of the total 639 BL Lacs analyses for 
variability, 304 sources show variability.  Similarly, out of the 481 FSRQs studied for 
flux variability, 371 are found to be variable. Thus, about 80\% of FSRQs are variable, while only 
about 50\% of BL Lacs are variable. We find mean $F_{var}$ values of 0.55 $\pm$ 0.33 and
0.47 $\pm$ 0.29 for FSRQs and BL Lacs, respectively. Thus FSRQs are more variable than BL Lacs in the
$\gamma$-ray band. This difference in the $\gamma$-ray flux variations between FSRQs and BL Lacs
is can be explained by the location of the inverse Compton peak in their broad-band SED with respect
to the {\it Fermi} observing band. Among different sub-classes of blazars, LSPs are more variable
followed by ISP and HSP blazars. The ensemble structure function analysis also shows that FSRQs are more 
variable than BL Lacs.
\item FSRQs show the highest DC of variability of 66\% relative to BL Lacs that show a 
DC of 36\%.
\item The majority of FSRQs and BL Lacs in our sample show time scales of variability of about 20 days.
This time scale could be related to processes in the jets of these sources. The distribution of timescales
between FSRQs and BL Lacs are different.
\item Statistically $F_{var}$ is not found to be not correlated with either M$_{BH}$ and $\delta$. Also, the time scale of the $\gamma$-ray flux variability does not show statistically significant correlation between
$M_{BH}$ and $\delta$.
\end{enumerate}

So our analysis to characterise the $\gamma$-ray flux variability on monthly-like time scales of the 1120 blazars for the period of nine years indicates that FSRQs are more variable than BLLACs, which is also explained by the analysis of the ensemble structure function and the duty cycle. And the time scale of variability and $F_{var}$ do not significantly correlate with M$_{BH}$ and $\delta$.

\begin{acknowledgements}
This publication makes use of data products from the {\it Fermi} Gamma-ray Space Telescope and
accessed from the Fermi Science Support Center\footnote{https://fermi.gsfc.nasa.gov/ssc/data/access/}
\end{acknowledgements}

\bibliographystyle{aa}
\bibliography{ref}
\end{document}